\documentclass[fleqn,usenatbib]{mnras}

\usepackage{newtxtext,newtxmath}

\usepackage[T1]{fontenc}
\usepackage{ae,aecompl}

\usepackage{graphicx}	
\usepackage{amsmath}	
\usepackage{amssymb}	

\newcommand{\flux}{\hbox{erg~cm$^{-2}$~s$^{-1}$}}

\title[Still No Star]{The Search for Failed Supernovae with the Large Binocular Telescope: N6946-BH1, Still No Star}

\author[C.~M. Basinger et al.]{
C.~M. Basinger$^{1}$,
C.~S. Kochanek$^{1,2}$, 
S.~M. Adams$^{3}$,
X. Dai$^{4}$,
K.~Z. Stanek$^{1,2}$
\\
$^{1}$Department of Astronomy, The Ohio State University, 140 West 18th Avenue, Columbus, OH 43210, USA \\
$^{2}$Center for Cosmology and AstroParticle Physics (CCAPP), The Ohio State University, 191 W. Woodruff Avenue, Columbus, OH 43210, USA \\
$^{3}$Cahill Center for Astrophysics, California Institute of Technology, Pasadena, CA 91125, USA \\
$^{4}$Department of Physics and Astronomy, University of Oklahoma, 440 W. Brooks Street, Norman, OK 73019, USA
}

\date{Accepted XXX. Received YYY; in original form ZZZ}

\pubyear{2020}

\begin{document}
\label{firstpage}
\pagerange{\pageref{firstpage}--\pageref{lastpage}}
\maketitle

\begin{abstract}
We present new Large Binocular Telescope, Hubble Space Telescope, and Spitzer Space Telescope data for the failed supernova candidate N6946-BH1. We also report an unsuccessful attempt to detect the candidate with Chandra. The $\sim$300,000$L_\odot$ red supergiant progenitor underwent an outburst in 2009 and has since disappeared in the optical. In the LBT data from May 2008 through October 2019, the upper limit on any increase in the R-band luminosity of the source is $2000 L_\odot$. 
HST and Spitzer observations show that the source continued to fade in the near-IR and mid-IR, fading by approximately a factor of 2 between October 2015 and September 2017 to 2900$L_\odot$ at H band ({\textit{F160W}}). Models of the spectral energy distribution are inconsistent with a surviving star obscured either by an ongoing wind or dust formed in the transient. The disappearance of N6946-BH1 remains consistent with a failed supernova.
\end{abstract}





\section{Introduction}

In modern theoretical models of supernovae (SNe), it is expected that some 10-30\% of core collapses fail to lead to an SNe and instead become black holes (e.g. \citealt{O'Connor2011}; \citealt{Pejcha2015}; \citealt{Sukhbold2016}) with a weaker intermediate transient 
(\citealt{Lovegrove2013}). Fall back SNe, where a successful SN explosion falls back onto the proto-neutron star and leads to black hole formation, are expected to be very rare \citep{Sukhbold2016}.
The existence of failed SNe naturally explains both the
apparent lack of higher mass progenitors to red supergiants (\citealt{Kochanek2008}; \citealt{Smartt2009}) and the compact object mass function (\citealt{Kochanek2014}; \citealt{Kochanek2015}).

From observations of interacting black hole binaries \citep{McClintock2006} and noninteracting binaries \citep{Thompson2019}, as well as LIGO (e.g., \citealt{Abbott2016}), we know that stellar mass black holes exist. 
However, the mass distribution of black holes inferred from interacting binaries and merging systems is intrinsically biased (e.g., \citealt{Belczynski2016}), and tells us nothing of the parent populations of stars which produced these black holes. 
There is basically no prospect of existing or next generation neutrino or gravitational wave detectors observing the formation of a new black hole (see \citealt{Adams2013}). This leaves surveys like that proposed by \cite{Kochanek2008} as the only current prospect of directly investigating the formation of black holes. 

In \cite{Kochanek2008}, we proposed a search for failed SNe by looking for massive stars which ``vanish''. We reported first results and a first candidate N6946-BH1 in \cite{Gerke2015}, with follow up observations of the candidate in \cite{Adams2017} and updated statistics in \cite{Adams2017b}. The implied
SN rates and the properties of the candidate were both consistent with expectations.

There is no doubt that the progenitor of N6946-BH1 was a massive luminous star that subsequently vanished in the optical and has at best a much fainter and fading near-IR counterpart. 
The absence of a warm-Spitzer counterpart implies that the star cannot be obscured by a present day, dusty wind. Newly forming dust is hot, leading to visible near/mid-IR emission unless the optical depths are very high.
This leaves the possibility of obscuration by dust formed during the transient associated with the vanishing of the star.

Ideally we would test this hypothesis with James Webb
Space Telescope (JWST) observations at 10-20$\mu$m to search for the mid-IR dust emission required by this hypothesis, 
and this remains a future prospect. 
However, time will also tell.
A required feature of any model obscuring the progenitor with an expanding shell of dust is that the star must eventually reappear. A uniform shell has an optical depth that drops as $t^{-2}$, and any evolution of the shell to be less homogeneous then accelerates the rate of decline \citep{Kochanek2012}. As time passes, increasingly implausible masses of ejecta and dust are also required to maintain the veil.

In \cite{Adams2017}, we reported on continued Large Binocular Telescope (LBT) observations of N6946-BH1 through the end of 2015 along with Hubble Space Telescope (HST) {\textit{F606W}} (V), {\textit{F814W}} (I), {\textit{F110W}} (J), and {\textit{F160W}} (H) band observations. In the LBT data we could see no evidence for any late-time brightening or fading of the source in the U, B, V, and R bands. In the optical HST bands, the star had effectively vanished, but there did seem to be a faint, fading near-IR counterpart whose luminosity was consistent with some models (e.g. \citealt{Perna2014}) 
for late-time accretion onto a newly formed black hole. 
While the rate of decline in the luminosity is consistent with the existence of an accretion disk, \cite{Fernandez2018} found it surprising that we observed the luminosity fading on these timescales, claiming that the bolometric luminosity should decay minimally or be roughly constant for many years.

In this paper, we extend the LBT monitoring observations through the end of 2019. We discuss new HST J and H band observations from September 2017 and a 4.5$\mu$m SST observation from September 2018, roughly two and three years after the previous observations, respectively. We still see no changes in the optical emission at the location of N6946-BH1, while the source faded in the near-IR and mid-IR.
We also report an
unsuccessful attempt to detect an X-ray counterpart with the Chandra X-ray Observatory. We discuss the data in \S2, and model various aspects of the data in \S3 and \S4. We discuss the results in \S5.
We adopt the revised distance to NGC~6946 of 7.7~Mpc
\citep{Anand2018} and a Galactic extinction of $E(B-V)=0.303$ based on the \cite{Schlafly2011} recalibration of \cite{Schlegel1998}.

\section{Observations and Data}

We obtained new HST WFPC3/IR
{\textit{F110W}} (J) and {\textit{F160W}} (H) band images of N6946-BH1 on 2017 September 15. We obtained 3 dithered images in both the J and H bands, with total integration times of 1350 and 1500 seconds, respectively. 
We reduced the data using {\sc dolphot} \citep{Dolphin2000}. 
We compare these images with HST Wide Field Camera 3 (WFC3) IR J and H band images from 2015 October 8, 
using {\sc iraf} to align the 2015 and 2017 images with a simple rotation and then using {\sc isis} (\citealt{Lupton1998}; \citealt{Alard2000}) to generate difference images between the two epochs.

We do photometry on the unaltered 2017 images using {\sc dolphot} with the parameters and procedures described in \cite{Adams2015}. 
We use a drizzled 2007 HST WFPC2 {\textit{F814W}} (I) band pre-outburst image as the astrometric reference as in \cite{Adams2017}. The 2017 images are aligned to the pre-outburst image with {\sc tweakreg} and {\sc tweakback} from the {\sc drizzlepac} package with an rms error of 0.05 arcsec. 
Using the pre-outburst image as the reference allows us to do photometry at the known location of the progenitor. Using the new H band image as the reference for photometry yields similar results. 
We will adopt the values obtained from the aligned images with the 2007 epoch as a reference to remain consistent with \cite{Adams2017}.

\begin{table}
\centering
\caption{Photometry.}
\begin{tabular}{c c c c c}
\multicolumn{1}{c}{MJD}
&\multicolumn{1}{c}{Date}
&\multicolumn{1}{c}{Filter}
&\multicolumn{1}{c}{Magnitude}
&\multicolumn{1}{c}{Telescope}\\
\hline
57303.3 & 2015-10-08 & UVIS $F814W$ & $26.02\pm0.16$ & \emph{HST} \\
57303.3 & 2015-10-08 & IR $F110W$ & $23.75\pm0.02$ & \emph{HST}\\
57303.3 & 2015-10-08 & IR $F160W$ & $22.38\pm0.02$ & \emph{HST}\\
57408.2 & 2016-01-21 & $3.6\>\mu\mathrm{m}$ & $18.47\pm0.18$ & \emph{SST}\\
57408.2 & 2016-01-21 & $4.5\>\mu\mathrm{m}$ & $17.46\pm0.09$ & \emph{SST}\\
58011.9 & 2017-09-15 & IR $F110W$ & $24.20\pm0.03$ & \emph{HST} \\
58011.9 & 2017-09-15 & IR $F160W$ & $22.89\pm0.04$ & \emph{HST} \\
58381.9 & 2018-09-20 & $4.5\>\mu\mathrm{m}$ & $18.01\pm0.15$ & \emph{SST}\\
\end{tabular}
\label{tbl:phot}
\end{table}

\begin{figure}
  \centering
	\includegraphics[width=\columnwidth]{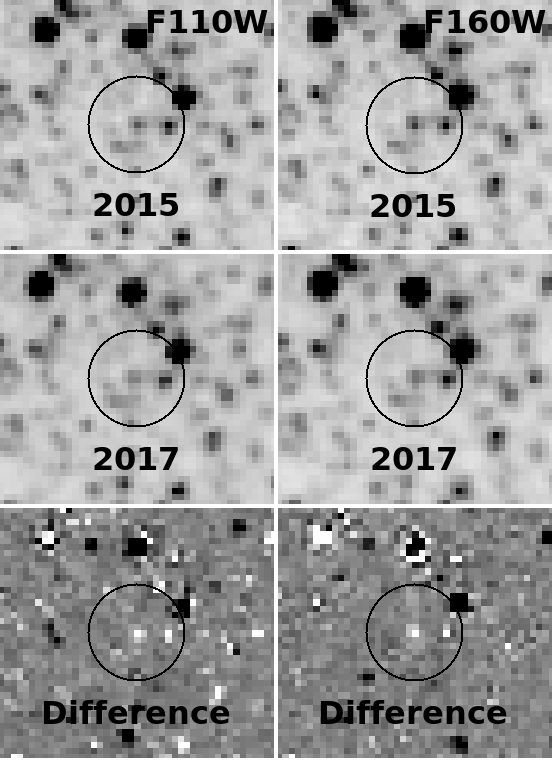}
    \caption{Near-IR HST images centered on the location of N6946-BH1. The top row shows the 2015 J (left) and H band (right) images. The middle row shows the corresponding 2017 images. The bottom row shows the difference images. The circles have a radius of 1 arcsec. Darker colors in all panels indicate a greater flux. The faint IR emission has faded by a factor of $\sim$2 from 2015 to 2017.}
    \label{fig:hst_new}
\end{figure}

Our 2015 and 2017 WFC3-IR J and H band images of the region surrounding the candidate are shown in Fig.~\ref{fig:hst_new}. We also show the difference between them, where black
(white) indicates a source that has become brighter (fainter). The source whose position is consistent
with that of the candidate appears to have faded between the two HST epochs by approximately 0.5~mag. 
The HST and SST magnitudes from observations since 2015 are shown in Table \ref{tbl:phot}. The old HST mags are repeated here from \cite{Adams2017}. Our rederived SST magnitudes for 2015 agree with the values reported in \cite{Adams2017}. 
As a luminosity ($\nu L_\nu$), the source has dropped from 
$2900L_\odot$ to $1900L_\odot$ at J band and from $4600 L_\odot$ to $2900L_\odot$ at H band. Shifting the luminosity of the progenitor from \cite{Adams2017} to the revised distance ($10^{5.29}L_\odot \rightarrow 10^{5.51}L_\odot$), these near-IR luminosities are less than 1\% of the luminosity of the progenitor
and correspond to luminosities of 5.8 and 8.8 $\times 10^{-3} L_E$ relative to the Eddington luminosity $L_E$ for a 10$M_\odot$ black hole.

\begin{figure*}
  \centering
	\includegraphics[width=\textwidth]{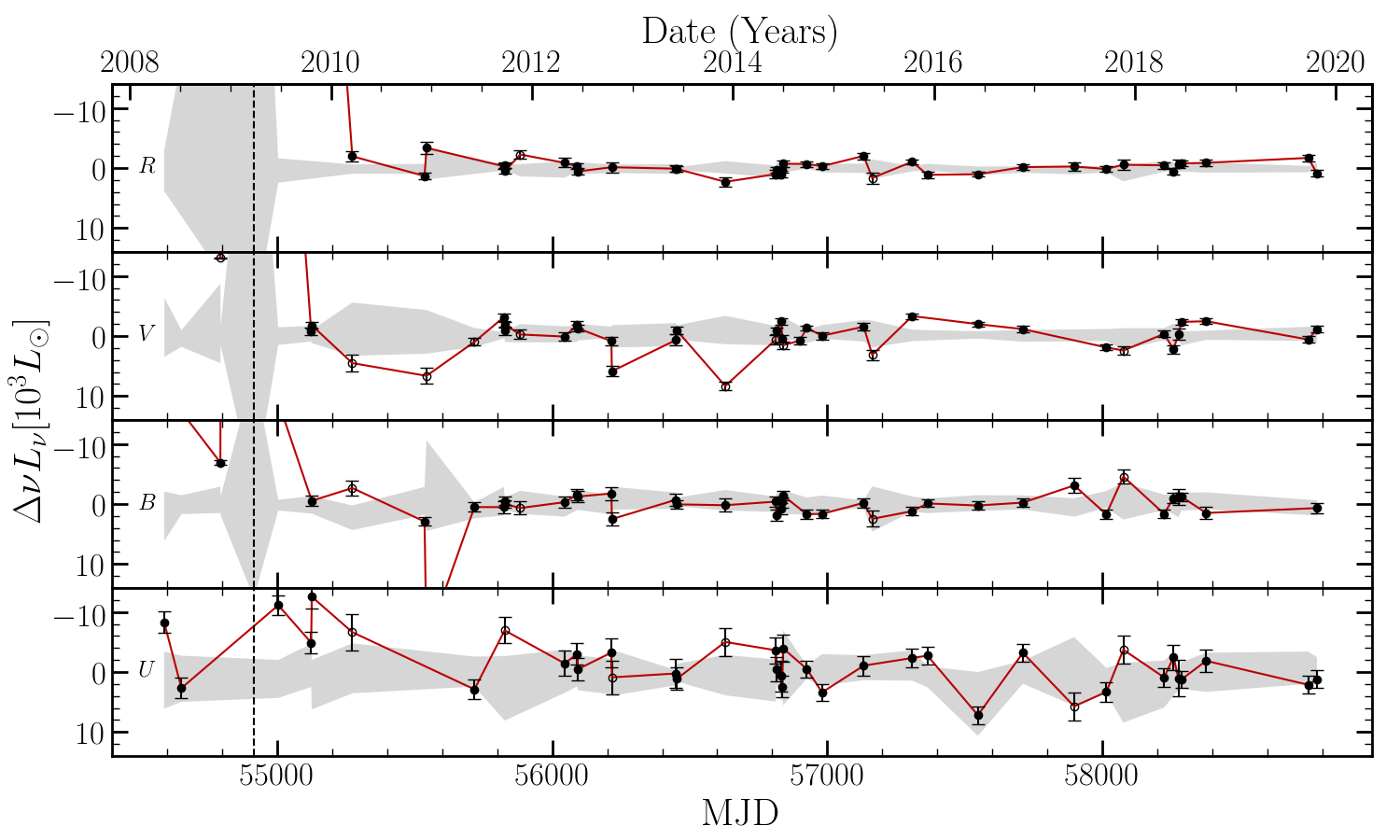}
    \caption{The UBVR difference photometry for N6946-BH1. Black circles indicate ``good'' data and open circles indicate ``low-quality'' data. The grey shaded region shows the dispersion of the comparison sample. The vertical dashed line indicates the observed transient peak (2009-03-25). The comparison sample is contaminated by the transient near its peak, leading to a large dispersion at these epochs.}
    \label{fig:lbt}
\end{figure*}

\begin{figure*}
  \centering
	\includegraphics[width=\textwidth]{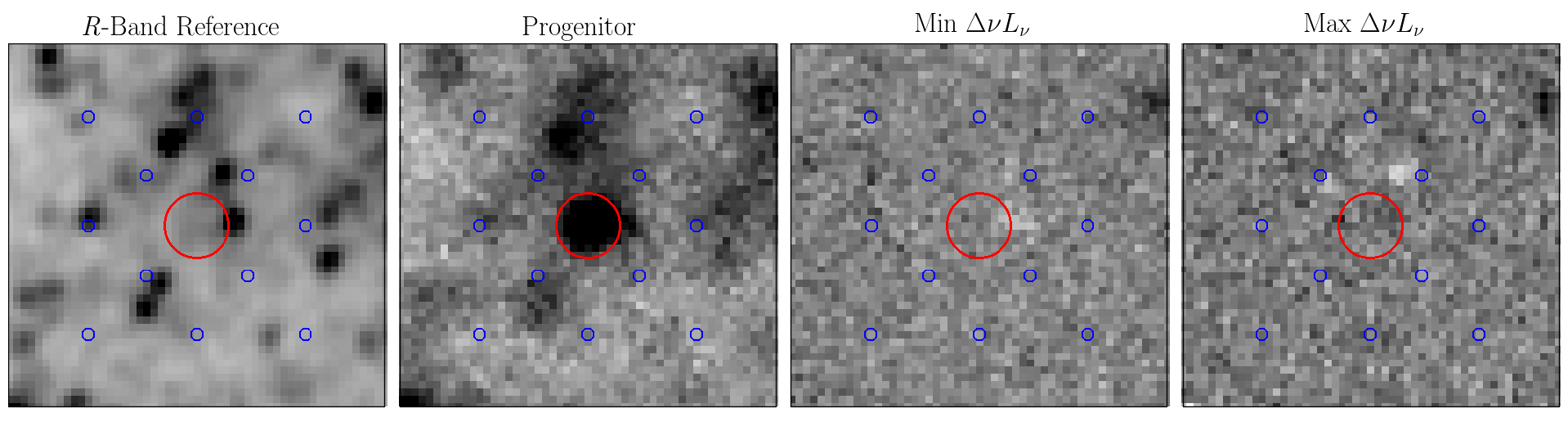}
    \caption{R-band images centered on the location of N6946-BH1. From left to right, the panels show the reference image, a pre-outburst image of the progenitor (2008-05-03), and two ``good'' post-outburst difference images: the minimum luminosity (2013-12-03), and the maximum luminosity (2015-04-20). The scales of the difference images are symmetric about zero. The red circles are 1 arcsec in radius and are centered on N6946-BH1. The smaller blue circles indicate the positions used for the comparison sample. In all panels, darker colors indicate a greater flux.}
    \label{fig:comparison}
\end{figure*}

\begin{figure*}
  \centering
	\includegraphics[width=\textwidth]{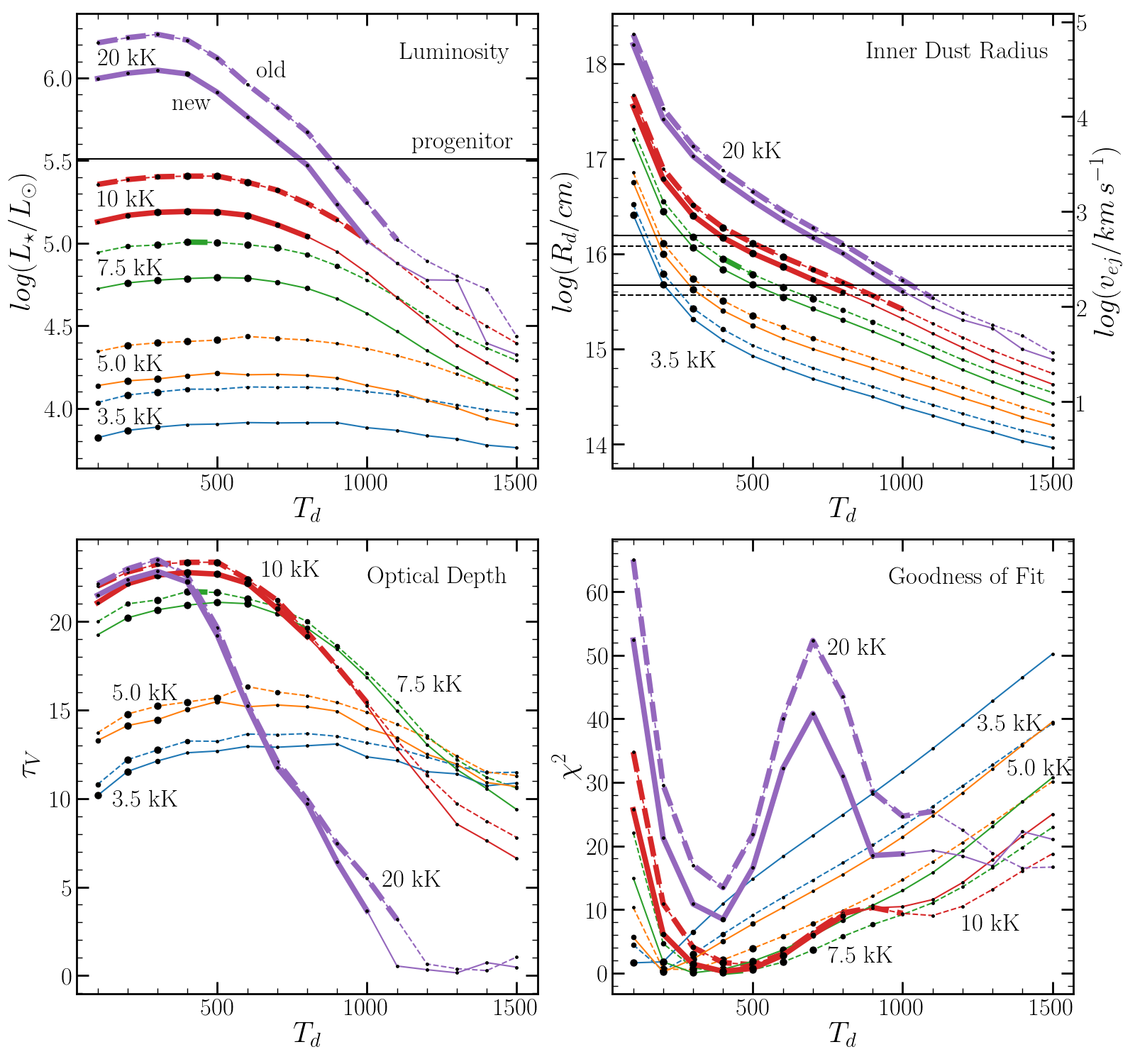}
    \caption{
    Parameters of the best fit models for a surviving source behind a dusty shell as a function of the dust temperatures ($T_{d}$, x-axis) at the inner edge of the shell. The upper left panel shows the best-fit luminosities of our models, the upper right panel shows the best-fit inner dust radii, the lower left panel shows the best-fit optical depths, and the lower right panel shows the corresponding $\chi^2$ values. Models for the old, 2015/2016 HST/SST measurements are shown with dashed lines, while models for the more recent 2017/2018 HST/SST measurements are shown by solid lines. Larger (smaller) black points indicate that the models fit better (worse). Each colored curve corresponds to a constant stellar temperature, $T_\star$: Blue - 3500 K, Yellow - 5000 K, Green - 7500 K, Red - 10000 K, and Purple - 20000 K. The thicker curves indicate models with a luminosity $L_\star > 10^5 L_\odot$. In the upper left, the solid black line shows the luminosity of the progenitor. In the upper right, the solid (dashed) black lines corresponding to the new (old) measurements show the $170 < v_{ej} < 560\text{ km s}^{-1}$ constraint on the ejecta velocity from \protect\cite{Adams2017}.}
    \label{fig:dusty}
\end{figure*}

We calculate the odds of an unrelated source being detected at the same location as the progenitor as in \cite{Adams2017}. The closest {\sc dolphot} source is 0.050 arcsec from the progenitor position. With a 4.2 $\text{arcsec}^{-2}$ surface density of sources, the odds of an unrelated source being detected at the same location purely by chance is 3.2\%. For sources as bright as or brighter than the detection in J band, the surface density is 0.76 $\text{arcsec}^{-2}$ and the likelihood is 0.6\%. For the detection in H band, the surface density is 0.64 $\text{arcsec}^{-2}$ and the likelihood is 0.5\%.

We incorporate new channel 2 (4.5$\mu$m) Spitzer Space Telescope (SST) data from 2018 September 20 (program ID: 13239; PI: K. Krafton) in addition to archival SST data into our analysis. 
We do aperture photometry following \cite{Adams2017}, using a 2.4 arcsec aperture and 2.4-4.8 arcsec radius sky annulus (Table \ref{tbl:phot}). We use the 3.6 and 4.5$\mu$m aperture corrections from Table 4.7 of the IRAC Instrument Handbook.\footnote{\url{https://irsa.ipac.caltech.edu/data/SPITZER/docs/irac/iracinstrumenthandbook/}}
We also use aperture photometry on the difference images to generate a light curve.
In this light curve, we observe that the most recent mid-IR flux is now below its minimum value in all previous epochs. 

The LBT data for NGC 6946 used here consists of 42 UBVR epochs taken from 2008 May 3 through 2019 October 24.
We used the {\sc iraf mscred} package for data reduction and {\sc isis} for photometry and generating light curves. 
The reference images used are identical to \cite{Adams2017} and they were generated using the best $\sim 20\%$ of the data from the first 6 years of the survey.

We extracted light curves both at the position of N6946-BH1 and for a comparison sample (Figs.~\ref{fig:lbt} and~\ref{fig:comparison}). Our comparison sample is a grid of 12 points surrounding N6946-BH1 with an inner grid spacing of 7 pixels and an outer grid spacing of 15 pixels as in \cite{Johnson2017}. The pixel scale is 0.2255 $\text{arcsec pixel}^{-1}$, so the comparison sample probes the space within several arcseconds of the candidate.
Three of the twelve points in the comparison sample were eliminated due to their proximity to variable stars. We flag ``low-quality'' data, defined by seeing $>$1\farcs5 or an {\sc isis} flux scaling factor $<$0.8, as in \cite{Johnson2017}. A low flux scaling factor indicates that the image was taken through clouds or at a high airmass.

We observed N6946-BH1 with ACIS-S3 \citep{Garmire2003} on board the Chandra X-ray Observatory \citep{Weisskopf2002} on 2016 September 28 with an exposure time of 58.6 ks.
We analyzed the level 2 event files from the standard pipeline products distributed from the Chandra X-ray Center, where the events were filtered using the standard ASCA grades of 0, 2, 3, 4, and 6, good flight time intervals, and status flags. 
We did not detect the X-ray counterpart of the failed supernova, and set 90\% confidence upper limits on the count rate as $5.1\times10^{-5}$ ct~s$^{-1}$ in the full 0.5-7 keV band, and 6.1, 3.9, and 3.9 $\times10^{-5}$ ct~s$^{-1}$ in the 0.5-1.2, 1.2-2, and 2-7 keV bands using the CIAO tool aprates.
Assuming a power law spectrum with a photon index of 1.7 and adopting $N_H = 3.9\times10^{21}$~cm$^{-2}$ for the combined Galactic and host galaxy absorption to X-ray sources in NGC 6946 based on the X-ray spectral fit results of \cite{Holt2003}, we obtain 90\% confidence unabsorbed flux upper limits of 8.1, 20.8, 2.9, and 7.1$\times10^{-16}\flux$, in the total, soft, medium, and hard bands, respectively.
Assuming a blackbody spectrum of 1~keV temperature yields similar limits on the flux of 7.4, 13.3, 2.8, and 6.7$\times10^{-16}\flux$.
Assuming a 10$M_\odot$ black hole and a distance of 7.7 Mpc, these upper limits correspond to 1500$L_\odot$ or $4.6\times10^{-3} L_E$ for the power law model and 1400$L_\odot$ or $4.2\times10^{-3} L_E$ for the blackbody.

If we assume that the black hole is radiating at the Eddington luminosity, we can estimate the neutral hydrogen column density, N(H), required to reduce the flux below our $5.1\times10^{-5}$ ct~s$^{-1}$ upper limit. 
Using the Chandra Proposal Planning Toolkit PIMMS v4.10,\footnote{\url{https://cxc.harvard.edu/toolkit/pimms.jsp}} we estimate N(H) $\gtrsim 1 \times10^{24} \text{ cm}^{-2}$.
This estimate does not strongly depend on whether we adopt a power law or blackbody model for the input flux.
The mass required to produce this absorption is roughly $M = N(H) 4 \pi R^2 m_H$ where R is the radius of the dusty shell and $m_H$ is the mass of hydrogen.
For R $\sim10^{16}$ cm (assuming an ejecta velocity of a few hundred $\text{km s}^{-1}$, see Fig.~\ref{fig:dusty}),
we find that $\gtrsim 1 M_\odot$ of material is required. This is small compared to the $13 M_\odot$ hydrogen envelope ejected in the failed SN model for a $\sim 25 M_\odot$ progenitor (\citealt{Woosley2002}), so the absence of X-rays is not surprising.
If there is accretion, the absence of X-rays rules out significantly higher ejecta velocities than expected for this scenario.

\begin{table}
\caption{Variability limits.}
\begin{tabular}{c c c c r c}
\multicolumn{1}{c}{Band}
&\multicolumn{3}{c}{Variability $[10^3\text{L}_\odot]$}
&\multicolumn{2}{c}{Slope [$10^3\text{L}_\odot$/yr]}\\
&\multicolumn{1}{c}{RMS}
&\multicolumn{1}{c}{$\langle\sigma^2\rangle^{1/2}$}
&\multicolumn{1}{c}{Sample}
&\multicolumn{1}{c}{Late-Time}
&\multicolumn{1}{c}{Sample}\\
\hline
$R$ & $0.9$ & $0.6$ & $0.8\pm0.1$ & $0.09\pm0.08$ & $0.04\pm0.03$\\ 
$V$ & $2.4$ & $0.6$ & $1.4\pm0.2$ & $0.18\pm0.21$ & $0.09\pm0.04$\\ 
$B$ & $1.6$ & $0.9$ & $1.6\pm0.4$ & $0.04\pm0.14$ & $0.10\pm0.09$\\ 
$U$ & $2.8$ & $2.0$ & $3.2\pm0.5$ & $-0.32\pm0.23$ & $0.20\pm0.15$\\ 
\end{tabular}
\label{tbl:var}
\end{table}

\section{Long-Term Variability and Trends}

Our difference imaging photometry for the candidate in the LBT UBVR images is shown in Fig.~\ref{fig:lbt}. 
The light curve of N6946-BH1 is in red and the range spanned by the light curves of the comparison sample is given by the grey shaded region. Good data points are indicated by black circles and ``low-quality'' data points are indicated by open circles. 
The R and U band references used for difference photometry were built without using any pre-outburst images which included the progenitor. However, the V and B band references did include a few such images, so we performed a correction on the V and B band light curves to account for this contamination. 
Aside from data points taken in poorer conditions, the variability seen at the location of the candidate is essentially indistinguishable from the variability at other nearby locations.

\begin{figure}
  \centering
	\includegraphics[width=\columnwidth]{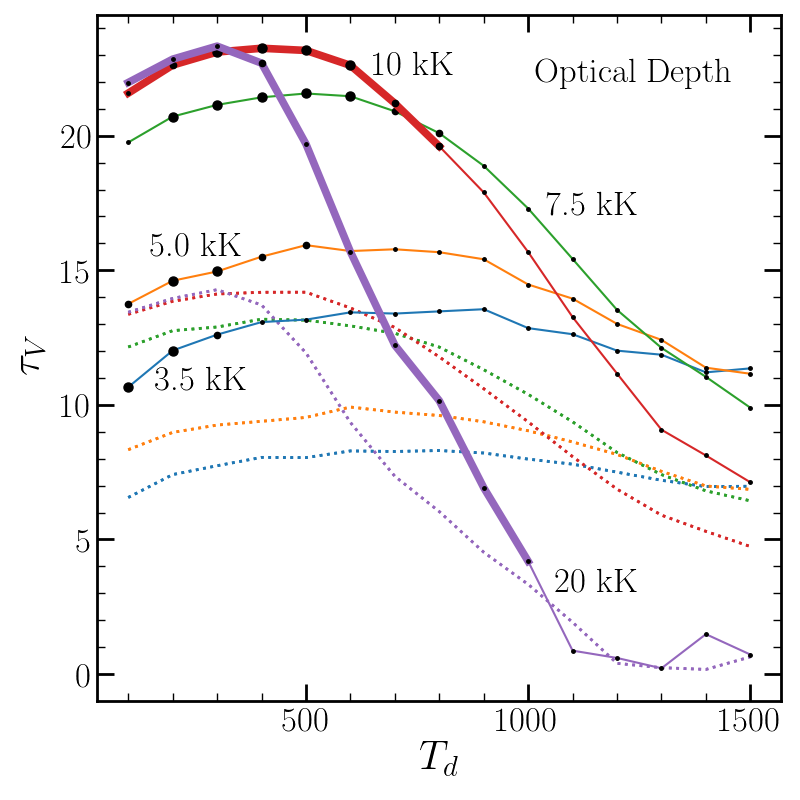}
    \caption{The corrected optical depths (solid lines) and expected optical depths (dotted lines). The corrected optical depth is defined as the optical depth required for the new measurements when the luminosity determined from the old measurements is held constant in our updated models. The expected optical depth follows a $t^{-2}$ evolution between the two epochs. The colors of the curves indicate the effective stellar temperature, the thick curves indicate a luminosity $L_\star > 10^5 L_\odot$, and the sizes of the black points roughly correspond to the goodness of fit as in Fig.~\ref{fig:dusty}.}
    \label{fig:tau}
\end{figure}

This is also illustrated in Fig.~\ref{fig:comparison} where we show the R-band reference image, a pre-outburst image of the progenitor, and R-band images with the minimum (2013-12-03) and maximum (2015-04-20) post-transient fluxes. The structure of the images at the position of the candidate are no different than that at the locations used to produce the comparison sample, again consistent with the lack of any variable source corresponding to the candidate.

Table \ref{tbl:var} characterizes the stochastic variability of N6946-BH1 using the root-mean-square (RMS) and the {\sc isis} noise estimate, $\langle\sigma^2\rangle^{1/2}$, of the post-outburst light curves. The average RMS for the comparison sample and the standard deviation about this average are reported under the first ``Sample'' column. The RMS tends to be about twice as large as the {\sc isis} noise estimate. Since the {\sc isis} noise estimate only considers Poisson statistics, this number is not surprising, and the comparison sample RMS is likely a better indicator of the limits on any potential variability. The RMS for N6946-BH1 is consistent with the RMS of the comparison sample except for the V band, which has a slightly inflated value due to the contribution from ``low-quality'' data points to the RMS (Fig.~\ref{fig:lbt}). We conclude that there is no evidence from the RMS for stochastic variability in the late-time light curve of N6946-BH1 at the level of $\sim1000 L_\odot$.

We also performed a linear fit, \( L(t) = At + B \), to the post-outburst light curves to look for long-term trends in luminosity. This used data from epochs beginning with 2012 April 28, ensuring that sufficient time had passed for the transient to fade. The slopes in Table \ref{tbl:var} are both positive and negative across the bands, and very close to 0 $\text{L}_\odot \text{yr}^{-1}$. 
From the start date until October 2019, the changes in luminosity are $700 \pm 600$, $1300 \pm 1600$, $300 \pm 1000$, and $-2400 \pm 1700 L_\odot$ in the R, V, B, and U bands (corresponding to 2$\sigma$ upper limits of 2000, 4500, 2400, and 1000$L_\odot$).
We also do linear fits to the points in the comparison sample and report the average absolute value of their slopes and their standard deviations in the second ``Sample'' column in Table \ref{tbl:var}. The variability of N6946-BH1 is consistent with the comparison sample, with a slope close to zero that is consistent with the absence of any long-term trends.

\begin{figure}
  \centering
	\includegraphics[width=\columnwidth]{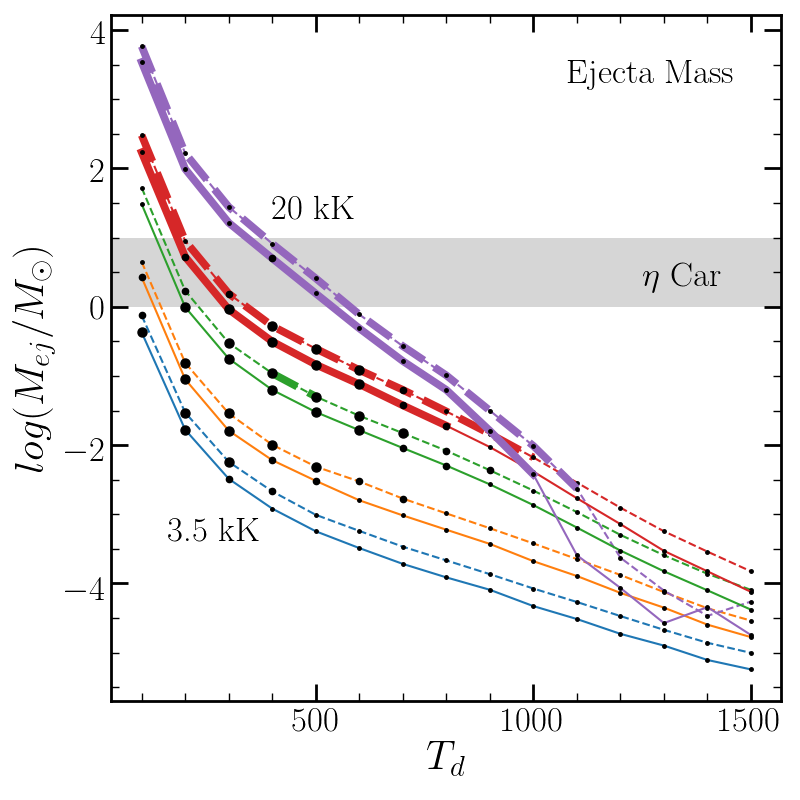}
    \caption{Ejecta mass $M_{ej} = 4{\pi}R_d^2{\tau}_V/{\kappa}_V$ implied by the optical depth for an opacity of $\kappa_V = 100 \text{ cm}^{2} \text{ g}^{-1}$. The grey shaded region shows a range of possible ejecta masses from $\eta$ Carinae's Great Eruption. The colors of the curves indicate the effective stellar temperature, the thick curves indicate a luminosity $L_\star > 10^5 L_\odot$, and the sizes of the black points roughly correspond to the goodness of fit as in Fig.~\ref{fig:dusty}.}
    \label{fig:mej}
\end{figure}

\section{SED Models}

If we adopt a range of temperature estimates for a surviving source and fit the new HST and SST near-IR and IR fluxes, we can determine the allowed luminosity of the source as a function of the temperature $T_d$ of any surrounding dust.
We do this using DUSTY (\citealt{Ivezic1997}; \citealt{Ivezic1999}; \citealt{Elitzur2001}), a radiative transfer code which models the light from a star surrounded by a spherical dusty shell or a dusty wind. As in \cite{Adams2017}, DUSTY is run inside an Markov Chain Monte Carlo (MCMC) wrapper to optimize the fits and to estimate uncertainties. Here, we only consider the case of a dusty shell as models with a dusty wind were previously ruled out in \cite{Adams2017}. Such models were too bright in the near/mid-IR due to the presence of hot dust.

For our models, we consider our 2015 HST I band image to be an upper limit, as there was no coincident source in the optical, but we consider both our 2015 and 2017 J and H band images to be detections of the source in the near-IR. The latest SST channel 1 and channel 2 data points (2016-01-21 and 2018-09-20) are among the lowest observed IR measurements and are treated as upper limits due to potential issues with aperture photometry in a crowded field given SST's resolution. 
We only consider the case of silicate dust, which is the type of dust favored to be formed from massive stars. 
The results will not strongly depend on whether we use silicate or graphitic dust.

We consider a range of stellar temperatures from $T_\star$ = 3500 to 20000 K. \cite{Adams2017} noted that a temperature of 3500 K was likely for the progenitor. At each stellar temperature, we run a model with a fixed inner edge dust temperature $T_{d}$, ranging from 100 to 1500 K. The outer and inner dust radii are fixed to a ratio of 2.0 as it has little effect on the fits.
With $T_\star$ and $T_d$ fixed, the fits determine the visual optical depth of the shell $\tau_V$ and the luminosity of the star $L_\star$. 
We show the results of our fits in Fig.~\ref{fig:dusty}. 
The results for the 2015 HST and 2016 SST data are shown by the dashed lines, and the results for the 2017 HST and 2018 SST measurements are given by the solid lines. The thicker lines indicate a luminosity $L_\star > 10^5 L_\odot$. The size of the black points roughly indicates the quality of the fits, with larger points corresponding to smaller $\chi^2$ values.

\begin{figure}
  \centering
	\includegraphics[width=\columnwidth]{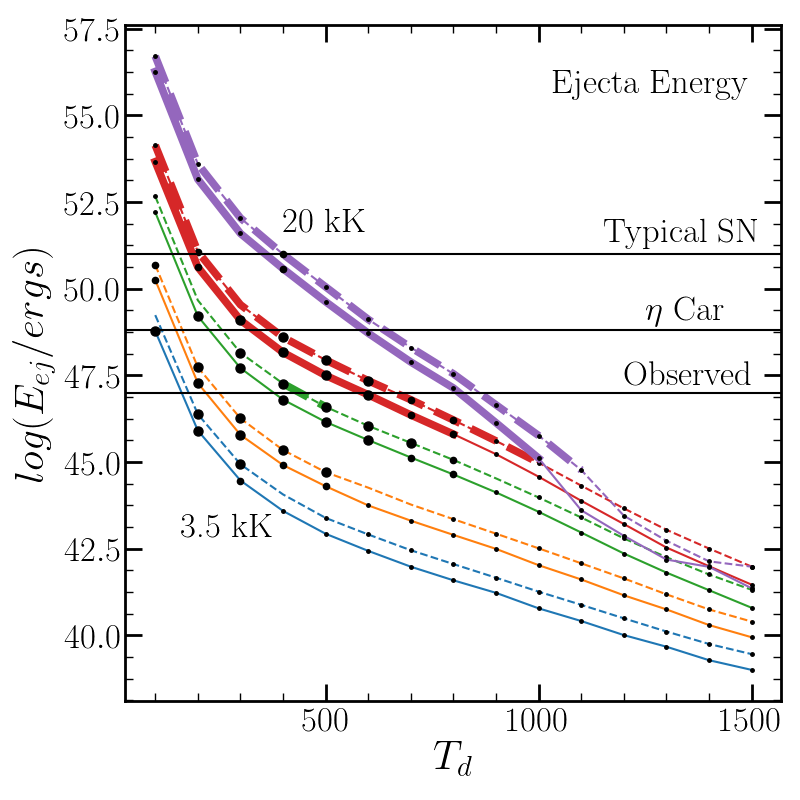}
    \caption{Energies required for the velocities and ejecta masses in Figs.~\ref{fig:dusty} and~\ref{fig:mej}. The black horizontal lines show a typical SN energy, the kinetic energy of $\eta$ Carinae's Great Eruption, and a rough estimate of the luminous energy of the observed transient. The colors of the curves indicate the effective stellar temperature, the thick curves indicate a luminosity $L_\star > 10^5 L_\odot$, and the sizes of the black points roughly correspond to the goodness of fit as in Fig.~\ref{fig:dusty}.}
    \label{fig:eej}
\end{figure}

In the upper-left of Fig.~\ref{fig:dusty} we see that only hot stars ($T_\star \gtrsim 10000$ K) can have present day luminosities similar to that of the progenitor ($\sim$300000 $L_{\odot}$, given by the black line). A star with a temperature similar to that of the progenitor ($T_\star \approx 3500$ K, the blue curve) must be far fainter. Because the near-IR emission has faded, the required temperature increases between the two HST epochs.

For ejecta velocities of $170 < v_{ej} < 560 \text{ km s}^{-1}$ (estimated by \citealt{Adams2017}), the ejecta lie between the black lines at roughly $R_d = 10^{16}$ cm. This allows the dust to be cool enough ($T_d \sim 800$ K) to be invisible to warm SST. Note that the hot star solution with $T_d \approx 800$ K fits badly because it struggles to keep the flux below the SST upper limits. Solutions with hotter dust require very slow expansion speeds and essentially approach a dusty wind model. Very cold dust generally requires very fast ejecta velocities and the large dust radii also require larger dust masses. Models with either very hot or very cold dust are generally bad fits.

In the lower left of Fig.~\ref{fig:dusty}, we see that the optical depths required to fit the new epochs are generally lower than those required to fit the SED in 2015/2016. If we adopt the epochs of the HST observations as the dates for the SEDs and 2008 Nov 25 as the date of the outburst, 
then the predicted change in the optical depth between the two observations for a $t^{-2}$ scaling is $\tau_{V_1} \approx 0.6 \text{ }\tau_{V_0}$ since the observations were made 6.9 and 8.8 years after the outburst. While the SED models do show small drops in the estimates of $\tau_V$, typically to a fraction $>$0.9 of $\tau_{V_0}$, the decrease is too small for an expanding shell over this time period. In fact, the observed drop is primarily driven by the lower luminosities found for the later epoch -- the drop in the near-IR flux leads to lower-luminosity solutions that need less dust to keep the SED below the optical and mid-IR flux limits. 

We can approximately correct for this effect by defining a luminosity-corrected optical depth for the second epoch of $\tau_{V_1,corr} \approx \tau_{V_1} + \ln(L_0/L_1$), where $L_0/L_1$ is the ratio of the luminosities between the two epochs. This correction is small, with an average ratio $\tau_{V_1,corr}/\tau_{V_1} = 1.08$. Fig.~\ref{fig:tau} compares the corrected optical depth estimates (dashed lines) with the prediction for an expanding shell extrapolated from our first epoch (dotted lines). Both with and without the correction, the optical depths at the second epoch are too large to be consistent with the expected evolution of the optical depth between the two epochs. 

The optical depth can be roughly converted to an ejecta mass by $M_{ej} = 4{\pi}R_d^2{\tau}_V/{\kappa}_V$ for a dust opacity scaled by $\kappa_V = 100 \text{ cm}^{2} \text{ g}^{-1}$. The ejecta mass is shown in Fig.~\ref{fig:mej}, with the black line representing the Great Eruption of $\eta$ Carinae for comparison (\citealt{Humphreys1994}; \citealt{Humphreys1999}). The newest measurements require a lower ejecta mass than the old measurements due to the decreasing $L_{\star}$ and $R_d$ in the models of the new measurements with lower IR fluxes.
Hiding the star requires an ejecta mass of $\sim 0.1-1 ({\kappa_V}/{100 \text{ cm}^{2} \text{ g}^{-1}})^{-1} M_\odot$ assuming a sufficiently hot star.
While the lack of optical depth evolution is inconsistent with an expanding shell, the mass budget is still plausible given that it is much less than the total hydrogen envelope mass of $\sim 13 M_\odot$ for a $\sim 25 M_\odot$ red supergiant.

Based on the velocities and ejecta masses in Figs.~\ref{fig:dusty} and~\ref{fig:mej}, we estimate the required ejecta energies with the results shown in Fig.~\ref{fig:eej}. 
The black horizontal lines show the energy of a typical SN ($\sim 10^{51}$ ergs), the kinetic energy of $\eta$ Carinae's Great Eruption ($10^{48.8}$ ergs, \citealt{Davidson1997}) and an estimate for the luminous energy of the transient ($\sim 10^{47}$ ergs). 
Plausible kinetic energies should be larger than the observed luminous energy.

For our calculations throughout this section, we computed the velocity, mass, and energy using the inner edge dust radius. 
If we instead consider the outer edge dust radius, we find a velocity and a mass that are two times larger, and an energy that is roughly five times larger. Such changes only strengthen our conclusions.

\section{Summary}
We present new LBT, HST, and SST data for our failed supernova candidate in NGC 6946. 
Investigation of the LBT light curve finds no evidence for either variability or long-term trends in the luminosity of any possible surviving star
with an upper limit of 2000$L_\odot$ for any re-brightening of the R-band luminosity. 
HST and SST observations in 2017/2018 show that the source has faded by nearly a factor of 2 in the near/mid-IR since the 2015/2016 epochs. 

Using our HST and SST observations, we create SED models for a potential surviving source surrounded by a spherical shell of dust.
The most interesting result from these models is the requirement for minimal evolution of the optical depth between the two epochs under the assumption of a surviving star, which does not match the behavior of an expanding dusty shell.
In addition, lower luminosity sources are more favored given the apparent decline in flux in the most recent HST and SST data. 
To hide a $\sim$300000$L_\odot$ star like the progenitor, our models require a much higher effective stellar temperature ($T_\star \gtrsim 10000$ K) for a surviving source. Otherwise, a star with similar effective temperature to the progenitor ($\sim$ 3500 K) must be intrinsically much fainter. Obscuration by an ongoing wind remains ruled out by the new data.

We also searched for X-ray emission with Chandra, which would be produced by fallback accretion onto a black hole. This unsuccessful attempt set an upper limit on the luminosity of an accreting, 10$M_\odot$ black hole at $4.2\times10^{-3} L_E$.
If such a black hole were radiating at the Eddington limit in X-rays, the observing limit would be met given an obscuring column of N(H) $\gtrsim 10^{24} \text{ cm}^{-2}$, roughly corresponding to an ejecta mass of $\gtrsim 1 M_\odot$ that would be easily exceeded by a failed supernova of a $\sim 25 M_\odot$ star with $13 M_\odot$ of ejecta.

We know of no other star which has been so dust enshrouded for so long. The most famous example which we can compare to our candidate is $\eta$ Carinae, which erupted and then dimmed over decade-long timescales (see \citealt{Humphreys1994}; \citealt{Humphreys1999}). Following its second outburst in the mid to late 1800s, its luminosity remained relatively constant at a few percent of its initial luminosity before gradually beginning to re-brighten in the mid 1900s.
However, there are two problems with this analogue as an explanation. First, $\eta$ Carinae would have been a tremendously luminous warm Spitzer source for this entire period, easily seen in any nearby galaxy like NGC 6946 (\citealt{Khan2010}; \citealt{Khan2013}).
Second, $\eta$ Carinae's evolutionary time scales, including its outburst were decades, while the outburst time scale of N6946-BH1 was less than a year. This would suggest that the ``recovery'' time scale for this system should also be far faster than observed for $\eta$ Carinae.

Our candidate remains at a luminosity of $\sim$1\% of its progenitor and we have not detected re-brightening nearly a decade after the initial outburst.
Continued optical and near-IR monitoring is one means of showing there is no surviving dust-obscured star.
It is, however, a somewhat indirect method.
After JWST launches, 10-20$\mu$m observations can directly test the possibility that a luminous star is hidden in cooler dust than can be detected with SST observations.
For now, N6946-BH1 remains an excellent failed supernova candidate with no compelling alternative explanation.

\section*{Acknowledgments}

We thank Rachel Patton for her assistance with {\sc dolphot} photometry. 
CSK and KZS are supported by NSF grants AST-1814440 and AST-1908952.
Based on observations made with the NASA/ESA Hubble Space Telescope, obtained from the Data Archive at the Space Telescope Science Institute, which is operated by the Association of Universities for Research in Astronomy, Inc., under NASA contract NAS5-26555. These observations are associated with programs GO-14266 and GO-15312.
This work is based in part on observations made with the Spitzer Space Telescope, which is operated by the Jet Propulsion Laboratory, California Institute of Technology under a contract with NASA.
The LBT is an international collaboration among institutions in the United States, Italy and Germany. LBT Corporation partners are: The University of Arizona on behalf of the Arizona university system; Istituto Nazionale di Astrofisica, Italy; LBT Beteiligungsgesellschaft, Germany, representing the Max-Planck Society, the Astrophysical Institute Potsdam, and Heidelberg University; The Ohio State University, and The Research Corporation, on behalf of The University of Notre Dame, University of Minnesota and University of Virginia.
Support for this work was provided by the National Aeronautics and Space Administration through Chandra Award Number 17500057 issued by the Chandra X-ray Center, which is operated by the Smithsonian Astrophysical Observatory for and on behalf of the National Aeronautics Space Administration under contract NAS8-03060.



\bsp	
\label{lastpage}

\begin{thebibliography}{}

\bibitem[Abbott et al.(2016)]{Abbott2016} Abbott B.~P., Abbott R., Abbott T.~D., Abernathy M.~R., Acernese F., Ackley K., Adams C., et al., 2016, PhRvL, 116, 061102
\bibitem[Adams et al.(2013)]{Adams2013} Adams S.~M., Kochanek C.~S., Beacom J.~F., Vagins M.~R., Stanek K.~Z., 2013, ApJ, 778, 164
\bibitem[Adams et al.(2017a)]{Adams2017} Adams S.~M., Kochanek C.~S., Gerke J.~R., Stanek K.~Z., Dai X., 2017, \mnras, 468, 4968
\bibitem[Adams et al.(2017b)]{Adams2017b} Adams S.~M., Kochanek C.~S., Gerke J.~R., Stanek K.~Z., 2017, MNRAS, 469, 1445
\bibitem[Adams \& Kochanek(2015)]{Adams2015} Adams S.~M., Kochanek C.~S., 2015, MNRAS, 452, 2195
\bibitem[Alard(2000)]{Alard2000} Alard, C.\ 2000, \aaps, 144, 363 
\bibitem[Alard \& Lupton(1998)]{Lupton1998} Alard, C., \& Lupton, R.~H.\ 1998, \apj, 503, 325 
\bibitem[Anand, Rizzi \& Tully(2018)]{Anand2018} Anand G.~S., Rizzi L., Tully R.~B., 2018, AJ, 156, 105
\bibitem[Belczynski et al.(2016)]{Belczynski2016} Belczynski K., Repetto S., Holz D.~E., O'Shaughnessy R., Bulik T., Berti E., Fryer C., et al., 2016, ApJ, 819, 108
\bibitem[Davidson \& Humphreys(1997)]{Davidson1997} Davidson K., Humphreys R.~M., 1997, ARA\&A, 35, 1
\bibitem[Dolphin(2000)]{Dolphin2000} Dolphin A.~E., 2000, PASP, 112, 1383
\bibitem[Elitzur \& Ivezi{\'c}(2001)]{Elitzur2001} Elitzur M., Ivezi{\'c} {\v{Z}}., 2001, MNRAS, 327, 403
\bibitem[Fern{\'a}ndez et al.(2018)]{Fernandez2018} Fern{\'a}ndez R., Quataert E., Kashiyama K., Coughlin E.~R., 2018, MNRAS, 476, 2366
\bibitem[Garmire et al.(2003)]{Garmire2003} Garmire G.~P., Bautz M.~W., Ford P.~G., Nousek J.~A., Ricker G.~R., 2003, SPIE, 4851, 28, SPIE.4851
\bibitem[Gerke, Kochanek \& Stanek(2015)]{Gerke2015} Gerke J.~R., Kochanek C.~S., Stanek K.~Z., 2015, MNRAS, 450, 3289
\bibitem[Holt et al.(2003)]{Holt2003} Holt S.~S., Schlegel E.~M., Hwang U., Petre R., 2003, ApJ, 588, 792
\bibitem[Humphreys \& Davidson(1994)]{Humphreys1994} Humphreys R.~M., Davidson K., 1994, PASP, 106, 1025
\bibitem[Humphreys, Davidson \& Smith(1999)]{Humphreys1999} Humphreys R.~M., Davidson K., Smith N., 1999, PASP, 111, 1124
\bibitem[Ivezic \& Elitzur(1997)]{Ivezic1997} Ivezic Z., Elitzur M., 1997, MNRAS, 287, 799
\bibitem[Ivezic, Nenkova \& Elitzur(1999)]{Ivezic1999} Ivezic Z., Nenkova M., Elitzur M., 1999, arXiv, astro-ph/9910475
\bibitem[Johnson, Kochanek \& Adams(2017)]{Johnson2017} Johnson S.~A., Kochanek C.~S., Adams S.~M., 2017, MNRAS, 472, 3115
\bibitem[Khan et al.(2010)]{Khan2010} Khan R., Stanek K.~Z., Prieto J.~L., Kochanek C.~S., Thompson T.~A., Beacom J.~F., 2010, ApJ, 715, 1094
\bibitem[Khan, Stanek \& Kochanek(2013)]{Khan2013} Khan R., Stanek K.~Z., Kochanek C.~S., 2013, ApJ, 767, 52
\bibitem[Kochanek et al.(2008)]{Kochanek2008} Kochanek C.~S., Beacom J.~F., Kistler M.~D., Prieto J.~L., Stanek K.~Z., Thompson T.~A., Y{\"u}ksel H., 2008, ApJ, 684, 1336
\bibitem[Kochanek, Szczygie{\l} \& Stanek(2012)]{Kochanek2012} Kochanek C.~S., Szczygie{\l} D.~M., Stanek K.~Z., 2012, ApJ, 758, 142
\bibitem[Kochanek(2014)]{Kochanek2014} Kochanek C.~S., 2014, ApJ, 785, 28
\bibitem[Kochanek(2015)]{Kochanek2015} Kochanek C.~S., 2015, MNRAS, 446, 1213
\bibitem[Lovegrove \& Woosley(2013)]{Lovegrove2013} Lovegrove E., Woosley S.~E., 2013, ApJ, 769, 109
\bibitem[McClintock \& Remillard(2006)]{McClintock2006} McClintock J.~E., Remillard R.~A., 2006, csxs.book, 157
\bibitem[O'Connor \& Ott(2011)]{O'Connor2011} O'Connor E., Ott C.~D., 2011, ApJ, 730, 70
\bibitem[Pejcha \& Thompson(2015)]{Pejcha2015} Pejcha O., Thompson T.~A., 2015, ApJ, 801, 90
\bibitem[Perna et al.(2014)]{Perna2014} Perna R., Duffell P., Cantiello M., MacFadyen A.~I., 2014, ApJ, 781, 119
\bibitem[Schlafly \& Finkbeiner(2011)]{Schlafly2011} Schlafly E.~F., Finkbeiner D.~P., 2011, ApJ, 737, 103
\bibitem[Schlegel, Finkbeiner \& Davis(1998)]{Schlegel1998} Schlegel D.~J., Finkbeiner D.~P., Davis M., 1998, ApJ, 500, 525
\bibitem[Smartt et al.(2009)]{Smartt2009} Smartt S.~J., Eldridge J.~J., Crockett R.~M., Maund J.~R., 2009, MNRAS, 395, 1409
\bibitem[Sukhbold et al.(2016)]{Sukhbold2016} Sukhbold T., Ertl T., Woosley S.~E., Brown J.~M., Janka H.-T., 2016, ApJ, 821, 38
\bibitem[Thompson et al.(2019)]{Thompson2019} Thompson T.~A., Kochanek C.~S., Stanek K.~Z., Badenes C., Post R.~S., Jayasinghe T., Latham D.~W., et al., 2019, Sci, 366, 637
\bibitem[Weisskopf et al.(2002)]{Weisskopf2002} Weisskopf M.~C., Brinkman B., Canizares C., Garmire G., Murray S., Van Speybroeck L.~P., 2002, PASP, 114, 1
\bibitem[Woosley, Heger \& Weaver(2002)]{Woosley2002} Woosley S.~E., Heger A., Weaver T.~A., 2002, RvMP, 74, 1015


\end{thebibliography}
\end{document}